# Analytical estimates of secular frequencies for binary star systems

Á. Bazsó and E. Pilat-Lohinger

University of Vienna, Department of Astrophysics, Türkenschanzstr. 17,
A-1180 Vienna, Austria
email: akos.bazso@univie.ac.at

**Abstract.** Binary and multiple star systems are extreme environments for the formation and long-term presence of extrasolar planets. Circumstellar planets are subject to gravitational perturbations from the distant companion star, and this interaction leads to a long-period precession of their orbits. We investigate analytical models that allow to quantify these perturbations and calculate the secular precession frequency in the dynamical model of the restricted three-body problem. These models are applied to test cases and we discuss some of their shortcomings. In addition, we introduce a modified Laplace-Lagrange model which allows to obtain better frequency estimates than the traditional model for large eccentricities of the perturber. We then generalize this model to any number of perturbers, and present an application to the four-body problem.

## 1 Introduction

### 1.1 Variety of extrasolar planets

Extrasolar planets (for short exoplanets) are planets that orbit stars other than our Sun. In the past 20 years exoplanets have been detected in a manifold of different orbital configurations.

It was a big surprise to detect massive planets with orbital periods as low as a few days. These objects, termed Hot Jupiters (HJs), have no matching counterpart in our solar system and their formation mechanisms are not well understood. Generally it is believed that such planets cannot form in-situ where they are observed now, but rather they have to form at larger distance to their host star and migrate inwards (Lin & Papaloizou, 1986). There must be some mechanism to stop this migration or else the planet would end up falling onto the star, but despite some models have been proposed (e.g. Ward, 1998; Masset & Papaloizou, 2003) there is no definite answer yet. Beaugé & Nesvorný (2012) studied the origin

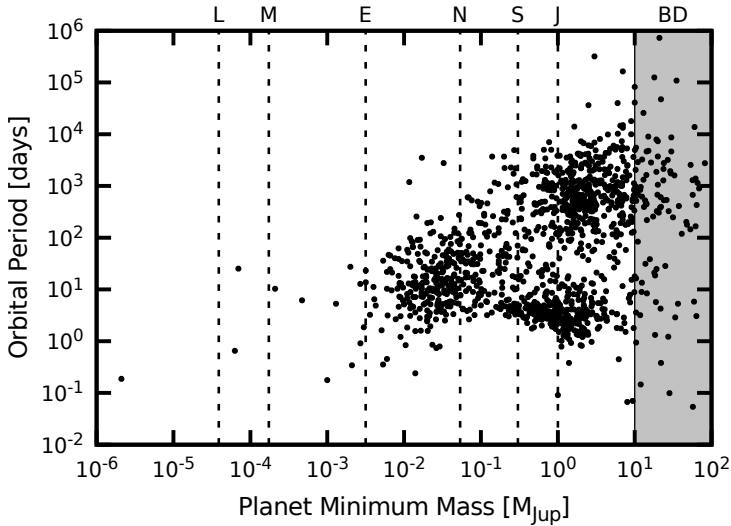

based on exoplanet.eu, 2017-01-25

**Figure 1.** Exoplanet minimum mass versus orbital period diagram. Objects located in the grey region are mainly brown dwarfs (BD). The letters on the top indicate the following planets: J = Jupiter, S = Saturn, N = Neptune, E = Earth, M = Mercury, L = Moon.

of HJs from chaotic large scale planet-planet scattering processes in multi-planetary systems. Another approach by Naoz et al. (2012) tried to explain the occurrence of HJs in binary stars by the Kozai mechanism and secular interactions with the companion star.

At the other extreme, exoplanets were also directly observed in very wide orbits far away from the host star (Marois et al., 2008). Also for this type of objects there are different competing formation mechanisms but none can completely explain these observations (Dodson-Robinson et al., 2009). The proposed formation mechanisms are:

1. in-situ formation in wide orbits by gravitational instability,

2. formation by core accretion in the inner parts of the disc and subsequent outward migration, or

3. scattering from the inner disc.

Among the key parameters for the classification of exoplanets are their masses and distances to the host star (equivalent to orbital periods). It is instructive to examine the minimum masses and periods of detected planets. Figure 1 presents an overview of the orbital period – planet mass ($M \sin i$) parameter plane for more than 1300 confirmed exoplanets from the Extrasolar Planet Encyclopaedia[1] (Schneider et al., 2011). As the planet's true masses are unknown in many cases because of the uncertain inclination of their orbital plane to the line of sight[2], we have to make do here with the minimum masses. Note that 'planets' with a minimum mass above 13 Jupiter masses ($M_J$) are generally considered to be brown dwarfs (indicated by the grey shaded area). The diagram is sparsely populated by exoplanets with masses lower than about $10^{-2}$ $M_J$, mainly because observational surveys in general have a bias towards high-mass objects with small orbital periods. This bias is present for both radial velocity surveys as well as transit surveys. Current radial velocity search programs have typical Doppler precisions of $1 - 5$ m s$^{-1}$ (Fischer et al., 2016), which is still insufficient to detect Earth-like planets in the habitable zone of Sun-like stars.[3] On the other hand transit surveys are susceptible to geometric effects that also lead to a bias (for details see Kipping & Sandford, 2016).

## 1.2 Exoplanets in multi-stellar systems

A considerable fraction of stars in the solar neighbourhood happen to be members of binary or multiple star systems. About 45 % of all Sun-like stars belong to multiple star systems as it was deduced from different observational surveys (Duquennoy & Mayor, 1991; Raghavan et al., 2010; Duchêne & Kraus, 2013; Tokovinin, 2014). In general, it is uncertain whether all of these systems really host exoplanets. For the known exoplanet host stars several studies (Raghavan et al., 2006; Mugrauer & Neuhäuser, 2009; Roell et al., 2012) published estimates on the stellar multiplicity, which yield the percentage of binary stars to be $10 - 15$ % among them, whereas multiple stars (with three or more components) amount to approximately 2 % of all exoplanet hosts.

The important point here is that exoplanets not only accompany single stars, but also binary, triple, and higher multiplicity stellar systems. Currently, there is an ongoing debate on whether or not multiple stars are a more hostile environment than single stars for the formation and long-term dynamical evolution of planets (Bromley & Kenyon, 2015; Kraus et al., 2016). Several studies have investigated the process of planet formation in the circumstellar discs of binary stars (see review by Thebault & Haghighipour, 2015), however they only arrived at the conclusion that the separation between the stars plays a major role. For stellar separations in the range $10 - 1000$ astronomical units (au) circumstellar exoplanets are about a factor of two less frequent than for a single star (Wang et al., 2014). Nevertheless, Jang-Condell (2015) has shown that even tight binaries, with separations of less than 30 au between the two stars, are able to form planets as massive as Jupiter from their truncated proto-planetary discs.

---

[1] http://exoplanet.eu/catalog/

[2] This is true especially for planets detected by radial velocity surveys (see discussion in Perryman, 2011).

[3] A planet of 1 (10) $M_\oplus$ at a typical distance of 1 au would produce a maximum RV signal of 0.09 (0.9) m s$^{-1}$ (Eggl et al., 2013).

From a dynamical point of view a binary star can host exoplanets in three different ways (see, e.g. Rabl & Dvorak, 1988; Dvorak et al., 1989):

1. The first possibility is a circumstellar planet (satellite or S-type) that orbits only one star, similar to an ordinary planet around a single star.

2. Another configuration is a circumbinary planet (planetary or P-type) that orbits the center of mass of both stars in a distance of a few times the stellar separation.

3. The third option is a Trojan planet (T-type) that moves in the vicinity of the Lagrangian libration point L4 or L5, similar to the Trojan asteroids of Jupiter.

However, in the last case the mass ratio of the binary is constrained to be lower than a critical value. This condition is not fulfilled by most Sun-like star pairs, but rather by star–brown dwarf pairs (Schwarz et al., 2015). For a study on Trojan planets in multi-planetary systems see Schwarz et al. (2017) in this volume.

Concerning the dynamical stability of these configurations there are countless studies on both real binary systems (including the long list of detected systems with the *CoRoT* and *Kepler* satellites) and purely theoretical model systems.

In this article we concentrate on circumstellar planets in binary star systems where the secondary star acts as a distant perturber. This condition is very important, because if we require Laplace stability[4] for the planet, then there exists a critical minimum distance ($a_c$) for the secondary star. The critical distance depends on the mass ratio of the two stars, their separation ($a_B$), and the eccentricity of both the secondary ($e_B$) and the planet ($e_P$). In the following we will discuss only a limited selection of articles and their results.

Rabl & Dvorak (1988) defined for the first time an empirical formula for the critical distance $a_c$ which they derived from numerical integrations of the elliptic restricted three-body problem. Their formula included only the dependence on the binary's eccentricity $e_B$, since they used an equal mass binary, and they assumed a circular orbit for the planet.

---

[4]According to Szebehely (1984) this means: All solutions stay in bounded regions of the phase space; no collisions or escapes may occur.

**Table 1.** Ratio of the semi-major axis of the secondary star ($a_B$) to that of the discovered planet ($a_P$) for various binary systems.

| System | $a_B$[au] | $a_P$ [au] | Ratio |
|---|---|---|---|
| HD 131399 Ab | 309 | 82 | 3.8 |
| HD 196885 Ab | 21 | 2.60 | 8.1 |
| HD 222404 Ab | 20 | 2.05 | 9.9 |
| Kepler-420 Ab | 5.3 | 0.38 | 13.9 |
| HD 41004 Ab | 23 | 1.64 | 14.0 |
| HD 126614 Ab | 36 | 2.35 | 15.4 |
| HD 19994 Ab | 100 | 1.42 | 70.4 |
| HD 177830 Ab | 97 | 1.22 | 79.4 |
| HD 1237 Ab | 68 | 0.49 | 138.8 |
| HD 13445 Ab | 19 | 0.11 | 172.7 |
| HD 177830 Ac | 97 | 0.51 | 188.8 |
| HD 128621 Bb | 23 | 0.04 | 575.0 |
| HD 120136 Ab | 45 | 0.05 | 978.3 |

Holman & Wiegert (1999) extended these older results by also investigating different mass ratios for a much longer integration time. They also provided an empirical formula that determines the boundary between regular and chaotic motion; however, the planet still started on a circular orbit in their study.

This restriction was removed when Pilat-Lohinger & Dvorak (2002) investigated the influence of a non-zero planetary eccentricity on the stability. They used the Fast Lyapunov Indicator (FLI) to identify $a_c$ as a function of the usual parameters mass ratio and $e_B$. According to their results the binary eccentricity is more effective for reducing the stable region, although a large planetary eccentricity also leads to a decrease of $a_c$.

In summary, the theoretical results for circumstellar planetary systems show that the typical critical distance of the secondary star is at least $2-4$ times the planet's distance from the host star, while in practice we find for observed binary stars that the secondary is between $4-1000$ times more distant than the planet. Table 1 shows the distance ratio $a_B/a_P$ for a selection of real binary star systems.[5] For more details on the system

---

[5]See the Catalogue of Exoplanets in Binary Star Systems of Schwarz et al. (2016) at `http://www.univie.ac.at/adg/schwarz/multiple.html`.

HD 131399 see Funk et al. (2017) in this volume, who investigate the dynamical stability of the exoplanet in this triple star system which has the most extreme distance ratio up to date.

## 1.3   Secular perturbations of planetary systems

One would naturally expect that the influence of the secondary star on any circumstellar planet is minimal in view of the large stellar separations for most real binaries. After all, the strength of Mean Motion Resonances (MMR) becomes weaker with increasing order of the resonance (Murray & Dermott, 1999), i.e. with larger distance between the two interacting bodies. However, there are also secular perturbations that can play an important role.

One example of such a case was presented by Pilat-Lohinger et al. (2016) who studied the binary system HD 41004 that consists of a K-type star with a Jupiter-like giant planet at $a_P = 1.64$ au and an M-dwarf companion at a distance of $a_B = 23$ au. The secondary induces a secular precession of the pericenter for the giant planet which in turn leads to a linear secular resonance (SR) in the region at $\approx 0.4$ au. This SR exists for both co-planar as well as inclined orbits and affects the region interior to the Habitable Zone (HZ) of the host star. But what would happen if the SR were exactly in the HZ?

This question was the main trigger for the survey of Bazsó et al. (2017). They investigated a sample of 11 binaries with circumstellar planets to determine in how many of them an SR affects the HZ. According to their results there are two cases where the SR is located inside the HZ, while in some more cases the SR is close to the HZ.

Both studies mentioned above made use of a semi-analytical method to investigate the secular dynamics of the binaries. The model treats the binary star – giant planet – test planet[6] restricted four-body problem as two coupled three-body problems. In a first step the giant planet's precession frequency is numerically determined by frequency analysis from the output of a single numerical integration. Then any analytical perturbation theory can be used to find the locations of linear SR; specifically, in the aforementioned papers the authors applied the Laplace-Lagrange theory (see section 2.2.5).

However, a combination of analytical models that preferably involve simple formulas for the giant planet's frequency (instead of numerically solving the equations of motion) would facilitate the study of a wide variety of binary configurations, and would also be useful for parameter studies of individual systems. The aim of this contribution is to review and compare various analytical models that can be used to calculate the planet's secular frequency.

---

[6]The 'test planet' represents a terrestrial planet, but it is treated as a massless test particle.

# 2 Analytical methods and models

## 2.1 Dynamical models

In the following sections we introduce step-by-step the dynamical models that were used for this work.

### 2.1.1 Full three-body problem

The binary star plus circumstellar planet configuration can be naturally considered as a hierarchical three-body problem, where the planet orbits the host star, while the secondary star plays the role of a distant perturber.

In the full three-body problem (3BP) all three bodies are massive ($m_i > 0, i = 1, 2, 3$), and there are no restrictions on their motion (i.e. full 18-dimensional phase space). For this most general case of the 3BP there is no complete solution except for special cases (Marchal, 1990).

We use the 3BP as the basis for our numerical simulations. These computations help to derive the reference values for the planet's secular frequency (for details see section 3).

### 2.1.2 Elliptic restricted three-body problem

In a first approximation we can assume that the exoplanet does not significantly influence the orbit of the secondary star. This assumption is reasonable in view of the large difference in the masses, which translates into a typical planet-star mass ratio of $10^{-3}$.

Hence, in this case we can use the elliptic restricted three-body problem (ER3BP) where the companion star would move on an unperturbed Keplerian orbit with fixed orbital elements ($a_B$, $e_B$), while the circumstellar planet is treated as a massless object and is subject to the gravitational perturbation of the secondary star (Szebehely, 1967). Moreover, we will assume that all objects move in a common plane, i.e. we only deal with co-planar systems.

We use the ER3BP as dynamical model for the analytical frequency estimates that will be introduced in section 2.2.

### 2.1.3    Restricted four-body problem

Our third at last dynamical model is the restricted four-body problem (R4BP). This consists of three massive bodies and one massless object. A possible application of the R4BP is a binary star system including a giant planet and an additional terrestrial planet, where the latter is treated as a massless object.

### 2.1.4    Definitions

Let us now introduce a few definitions that we will use throughout the rest of this article. Our definition of the stellar mass ratio is $\mu = m_B/m_A$, instead of the 'usual' definition of $\mu = m_B/(m_A + m_B)$. Here $m_A$ ($m_B$) is the mass of the host star (secondary star). This definition does not exclude the possibility that the secondary is more massive than the host star ($\mu > 1$). The variable $\alpha = a_P/a_B$ denotes the semi-major axis ratio of the planet ($a_P$) to the secondary star ($a_B$). Note that the system must be of hierarchical nature in order to be stable as discussed above, which results in $\alpha \ll 1$.

## 2.2    Listing of analytical models

In this section we introduce the different models and describe each of them briefly.

### 2.2.1    Heppenheimer model (HEP)

The Heppenheimer model was introduced to study the formation of planets in binary star systems (Heppenheimer, 1978). It is a restricted three-body problem including two massive stars besides a massless planet. The model is of first order in the masses, and second degree in the planet's eccentricity ($\mathcal{O}(e_P^2)$), while it reflects the exact functional dependence on the perturber's eccentricity ($e_B$). Further restrictions are that it only works for an external perturber (i.e. only for S-type binaries), and co-planar orbits (no inclination).

    The complete formula for the planet's secular frequency is given by

$$g_{Hep} = \frac{3}{4}\mu\alpha^3 n \left(1 - e_B^2\right)^{-3/2}. \tag{1}$$

The variable $n$ denotes the planet's mean motion around the host star, $n^2 = Gm_A/a_P^3$. We use the astronomical system of units, in which $G$ is expressed by the Gaussian gravitational constant ($k = 0.01720209895$) in units of au$^3\,M_\odot^{-1}\,$d$^{-2}$. Since $n$ is the only dimensional quantity in equation (1), $g$ will have the same units. For convenience in the applications we convert $g$ from radian per day to arcseconds per year, and will consistently use these units throughout the rest of this work. Note that equation (1) is independent of $e_P$, which means that it is only accurate for low planetary eccentricities.[7]

### 2.2.2 Giuppone model (GIU)

Giuppone et al. (2011) constructed a second order theory in the masses for the secular dynamics of planetesimals in binary star systems. Since these planetesimals emerge from the circumstellar disc they assumed a coplanar problem. Their aim was to find improved expressions for the forced eccentricities and secular frequencies. In their analysis they demonstrated that the second order model leads to significant improvements for these two parameters, however the full model is much too complex and unwieldy for everyday use.

To overcome this problem they made the following ansatz: Starting from the HEP model they introduced a second order correction term which was adjusted by multivariate linear regression to their full theory. The resulting expression for the GIU model is

$$g_{Giu} = g_{Hep} \left[ 1 + 32\mu\alpha^2(1 - e_B^2)^{-5} \right] \tag{2}$$

with $g_{Hep}$ as in equation (1). Note that the exponents of the parameters $(\mu, \alpha, e_B)$ were chosen by trial and error and are by no means derived from a theoretical perspective.

There is one important drawback of this model, namely that it was constructed and applied exclusively to the binary system $\gamma$ Cephei. Although the GIU model was not intended to be used for other cases, we will include it to the comparison.

### 2.2.3 Andrade-Ines & Eggl model (AND)

Andrade-Ines & Eggl (2017) established an empirical correction to the HEP model in the spirit of Giuppone et al. (2011), with the same goal to improve the quality of the secular frequency and forced eccentricity estimates. This semi-analytical correction has thus the same limitations as that of the original HEP model.

Starting from direct numerical simulations of binary stars they used a harmonic decomposition of the time-evolution of a suitable secular variable to evaluate the reference value for the frequency. Following this step they used a polynomial function in the variables eccentricity, mass and semi-major axis ratio to adjust the correction terms for the analytical model:

$$g_{And} = g_{Hep} \left(1 - \delta_g\right) \tag{3}$$

$$\delta_g = \sum_{k=1}^{M} A_k \alpha^{p_k} e_B^{q_k} \mu^{r_k} \tag{4}$$

with $g_{Hep}$ as in equation (1).

---

[7]Tests performed by the authors and practical experience suggest an empirical limit of $e_P \leq 0.2$; above this value the formula is still applicable but the neglected higher order terms in $e_P$ start to deteriorate the accuracy.

The sum contains a number of $M \leq 20$ different terms; the exponents take the values $p_k \in \{\frac{3}{2}, \frac{9}{2}\}$, $q_k \in \{0, 2, 4\}$, and $r_k \in \{\frac{1}{2}, 1, 2\}$. Note that unlike for the GIU model, in this case the exponents were chosen with reference to the theoretical work of Andrade-Ines et al. (2016).

### 2.2.4   Georgakarakos model (GEO)

In a series of papers Georgakarakos (2002, 2003, 2009) worked out a model for the eccentricity evolution (on short and secular time-scales) in hierarchical three-body systems. This model is applicable to both stellar and planetary mass objects (Georgakarakos, 2006), hence in principle it includes the dependence on all three masses. In the following we will assume a restricted problem (the planet's mass set to zero), though, to provide a fair comparison between the different models.

$$g_{Geo} = g_{Hep} \left[1 + \frac{25}{8} \frac{\mu}{\sqrt{1+\mu}} \alpha^{3/2} \frac{3 + 2e_B^2}{(1 - e_B^2)^{3/2}}\right] \tag{5}$$

with $g_{Hep}$ as in equation (1). This model is also of second order in the masses, as is visible from the square-root term in the brackets. Note that equations (1)–(5) all have the same common factor $g_{Hep}$, but they deviate in important details concerning their dependence on $(\mu, \alpha, e_B)$. Especially, AND and GEO include terms $\mathcal{O}(\alpha^{3/2})$, while GIU has $\mathcal{O}(\alpha^2)$.

### 2.2.5   Laplace-Lagrange model (LL)

The Laplace-Lagrange secular perturbation theory (LL) is a first order model with respect to the masses, and includes second degree terms in the eccentricity and inclination while higher degree terms are neglected (see Brouwer & Clemence, 1961; Murray & Dermott, 1999). It was developed as a first approximation to the dynamics of planets in the solar system, which explains the limitations listed above.

In the limit of a single external perturber (besides the host star) the proper secular frequency $g$ of the planet is calculated from

$$g_{LL} = \frac{1}{4}\mu\alpha^2 n b_{3/2}^{(1)}(\alpha), \tag{6}$$

where the expression $b_{3/2}^{(1)}$ is a special case of a Laplace coefficient (Murray & Dermott, 1999). The general form of a Laplace coefficient is defined by

$$\frac{1}{2}b_n^{(k)}(\alpha) = \frac{1}{2\pi}\int_0^{2\pi}\frac{\cos(k\varphi)}{(1 - 2\alpha\cos\varphi + \alpha^2)^n}\mathrm{d}\varphi. \tag{7}$$

This coefficient is an explicit function of the parameter $\alpha$, while the indices $n = 3/2$ and $k = 1$ are fixed in equation (6).

The integral in equation (7) does not admit a closed form solution, but rather it has to be evaluated either numerically, by means of hypergeometric functions (Brouwer & Clemence, 1961), or by developing it into a Taylor series in $\alpha$ which is convergent provided that $\alpha < 1$. Expanding the Laplace coefficient into a Taylor series we obtain

$$b_{3/2}^{(1)}(\alpha) = 3\alpha + \frac{45}{8}\alpha^3 + \frac{525}{64}\alpha^5 + \mathcal{O}(\alpha^7). \tag{8}$$

## 3   Application of the models

After this review of the analytical models we are ready to apply them to a couple of test cases. For this application we chose to use a synthetic binary star system consisting of two equal mass G-type stars of $1\,M_\odot$ each ($\mu = 1$), with a single planet of $1\,M_J$. We are interested in how well the models can estimate the secular frequency when the secondary's eccentricity varies between $0 \leq e_B \leq 0.6$.

Figure 2 shows a comparison of the analytical models from section 2.2 (using the ER3BP) to reference values obtained from numerical integrations. In these simulations the full three-body problem is solved numerically by means of the Lie-series method (Hanslmeier & Dvorak, 1984), as well as the Radau method (Everhart, 1974) from the Mercury package (Chambers, 1999). Eggl & Dvorak (2010) reviewed the characteristic properties of these methods and compared their efficiency. The reference frequency is obtained by Fourier analysis from the time series of the complex variable $z(t) = h(t) + ik(t)$, where $i = \sqrt{-1}$ and

$$h(t) = e\sin\varpi$$
$$k(t) = e\cos\varpi.$$

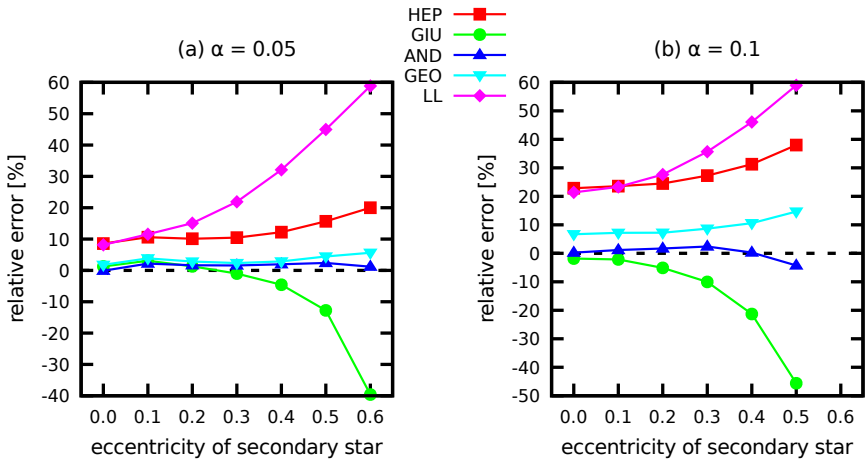

**Figure 2.** Comparison of analytical models (using the ER3BP) to the reference value derived from numerical simulation (using the full 3BP). Left panel (a): an equal mass binary with $a_B = 60$ au with a Jupiter mass planet at $a_P = 3$ au. Right panel (b): same as before with $a_B = 70$ au and $a_P = 7$ au.

For the Fourier part we applied as complementary tools the Fast Fourier Transform package FFTW[8] of Frigo & Johnson (2005) and the Frequency Modified Fourier Transform (FMFT) of Šidlichovský & Nesvorný (1996).

In the left panel (a) of Figure 2 the planet always has a semi-major axis of $a_P = 3$ au and starts on a circular orbit, while the binary has a separation of $a_B = 60$ au and starts at periastron. The dashed horizontal line marks the reference numerical frequency. We can observe that all methods deviate more or less from it, and most of them overestimate the frequency (i.e. they lie above the line). The LL model performs worst since it completely lacks the eccentricity dependence. It predicts a constant value for the frequency for all values of $e_B$ which leads to the strong increase in the relative error. The HEP model is somewhat more accurate with relative errors ranging in between $10 - 20$ %. The GEO and AND models are on average the best, with a slight advantage for the latter. The behaviour of the GIU model is remarkable in that it is among the best models for low eccentricities, then it exactly matches the numerical value (between $e_B = 0.2 - 0.3$), but finally strongly diverges for large eccentricities.

The other test case in panel (b) of Figure 2 is for a planet located at $a_P = 7$ au with the secondary star at $a_B = 70$ au. For this combination of system parameters the planet's orbit becomes chaotic for $e_B = 0.6$, so we do not include this data point in the plot. Basically the picture is the same as before, the only difference is that the relative errors are higher due to the larger value of $\alpha$. This affects mainly the HEP model which does not include higher order terms in $\alpha$ and so doubles its relative error. For large eccentricity the AND model performs quite well, although we can observe that it over-corrects and would overshoot like the GIU model for still higher $e_B$.

---

[8]See homepage at http://fftw.org/.

## 4   An extension of the LL model

The results from Figure 2 demonstrate that the LL model can potentially be as accurate as the HEP model, but only for low eccentricities $e_B \leq 0.1$. In fact, when combining equations (6) and (8), one gets

$$g_{LL} = \frac{1}{4}\mu\alpha^2 n \left(3\alpha + \mathcal{O}(\alpha^3)\right) = \frac{3}{4}\mu\alpha^3 n. \qquad (9)$$

This equation is actually very similar to equation (1), except for the missing term $(1 - e_B^2)^{-3/2}$.

A natural extension of the LL model would be to add this term as a 'correction' factor, i.e. to ad hoc introduce the eccentricity dependence into the model. Thus we define the *modified Laplace-Lagrange model* (LLM) by

$$g_{LLM} = \frac{1}{4}\mu\alpha^2 n b_{3/2}^{(1)}(\alpha)(1 - e_B^2)^{-3/2}. \qquad (10)$$

The basic structure of the formula remains the same; it is identical to the conventional LL model in the limit of $e_B \to 0$, while it is identical to the HEP model for $\alpha \to 0$. Thus, we can think of equation (10) as an interpolating formula between the LL and HEP models.

Unlike all other models from section 2.2 the LL model is applicable also in cases when there is more than one massive perturber. As mentioned before, the other models work only for a single external perturber (ER3BP). The one exception to this is the GEO model that indeed works also for a full 3BP. In the LL model the secular frequency of a massless test particle is just the sum of all individual contributions by all perturbing masses (see Murray & Dermott, 1999, chapter 7). Using this property in combination with the LLM we obtain

$$g = \frac{1}{4}n \sum_{j=1}^{N} \mu_j \alpha_j^2 b_{3/2}^{(1)}(\alpha_j) \left(1 - e_j^2\right)^{-3/2}. \qquad (11)$$

As a test case for $N = 2$ we have a binary star system with a giant planet and one or perhaps several (terrestrial) planets (all of them considered to be massless particles, R4BP). From Table 1 we chose the system HD 41004 as a representative (see Pilat-Lohinger et al., 2016). It consists of a $0.7\,M_\odot$ host star, a $0.4\,M_\odot$ secondary at $a_B = 23$ au, and a planet of $2.54\,M_J$ at $a_P = 1.64$ au. The eccentricities for the planet and secondary star are both set to 0.2.

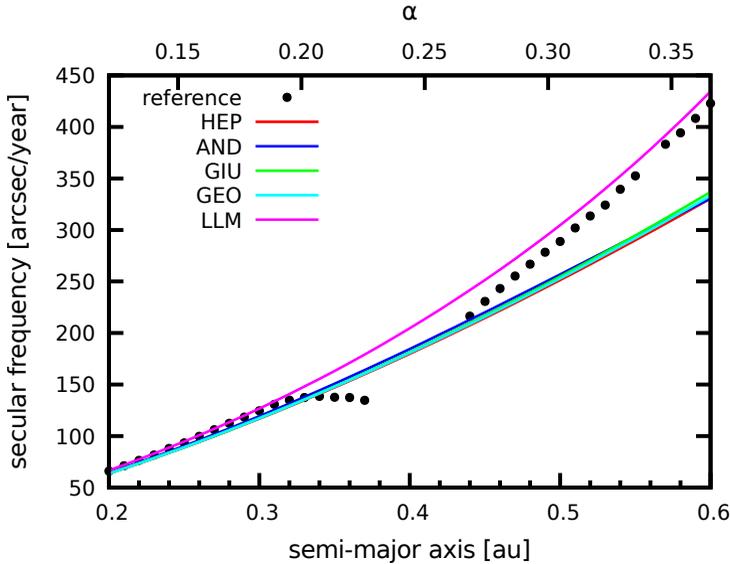

**Figure 3.** Comparison of analytical models to reference values obtained from numerical simulations for the system HD 41004. The dynamical model is the R4BP for LLM and the reference, while it is the ER3BP for all other models (HEP, AND, GIU, and GEO).

Figure 3 shows the comparison of the LLM (R4BP) to the other models (ER3BP). The data points in black represent the reference values, which were obtained from direct numerical integration of the R4BP. There are two gaps in this data set, one at $a \approx 0.4$ au and another at $a = 0.56$ au. The first one is associated to a linear SR with the giant planet which causes the numerical frequencies to diverge in the vicinity of this resonance. The smaller second gap is due to the 5:1 MMR with the planet. Only the LLM (pink curve) fits nicely to the numerical curve, although there is a small offset for larger $\alpha = a/a_P$ which means that the frequency is systematically overestimated.

All three-body models (HEP, GIU, AND, GEO) are essentially overlapping and follow the same trend. They agree quite well for small $\alpha < 0.2$, but afterwards they are unable to predict the correct frequency. The reason for this behaviour is that they include the contribution from the giant planet, but they fail to model the effect of the secondary star. In this re-spect, only the LLM is able to give quantitatively correct estimates for the test particle's secular frequency.

## 5 Conclusions

We provide a review of several analytical models that can be used to calculate secular frequencies for objects in the ER3BP. As an application we focus on hierarchical three-body problems such as circumstellar planets in binary star systems. We perform a comparison of these models and confirm that the Laplace-Lagrange (LL) model has a limited accuracy, because it does not take into account the perturber's eccentricity. As a workaround we introduce the modified LL model which contains a correction term for the perturber's eccentricity. This new model is then tested against the others for a R4BP, which demonstrates that it works and gives quantitatively correct results when compared to numerical data.

**Acknowledgements:** We thank the organizers of the workshop in Ammouliani as well as the editors of this proceedings volume. The authors acknowledge support by FWF projects S11608-N16 and P22603-N16.